\documentclass[
    ,final            
  ]
  {aipproc}

\layoutstyle{8x11double}

\usepackage{amsmath}
\usepackage{amsfonts}
\usepackage{amssymb}

\usepackage{epsfig}

\newcommand{\be}{\begin{equation}}
\newcommand{\ee}{\end{equation}}
\newcommand{\bea}{\begin{eqnarray}}
\newcommand{\eea}{\end{eqnarray}}

\newcommand{\GeV}{~\mathrm{GeV}}
\newcommand{\TeV}{~\mathrm{TeV}}

\begin{document}


\title{$B-L$ mediated SUSY breaking with radiative $B-L$ symmetry breaking}


\author{Tatsuru Kikuchi}
{
  address={Theory Division, KEK, Oho 1-1, Tsukuba, Ibaraki, 305-0801, Japan},
  email={tatsuru@post.kek.jp}
}

\author{Takayuki Kubo}
{
  address={Theory Division, KEK, Oho 1-1, Tsukuba, Ibaraki, 305-0801, Japan},
  altaddress={The Graduate University for Advanced Studies, Oho 1-1, Tsukuba, Ibaraki, 305-0801, Japan},
  email={kubotaka@post.kek.jp}
}

\begin{abstract}
We explore a mechanism of radiative $B-L$ symmetry breaking
in analogous to the radiative electroweak symmetry breaking.
The breaking scale of $B-L$ symmetry is related to the neutrino masses
through the see-saw mechanism.
Once we incorporate the ${\rm U}(1)_{\rm B-L}$ gauge symmetry in SUSY models,
the ${\rm U}(1)_{\rm B-L}$ gaugino, $\tilde{Z}_{B-L}$ appears, and it can mediate 
the SUSY breaking (Z-prime mediated SUSY breaking) at around the scale of $10^6$ GeV.
Then we find a links between the neutrino mass (more precisly the see-saw or $B-L$ scale
of order $10^{6}$ GeV) and the Z-prime mediated SUSY breaking scale.
It is also very interesting that the gluino at the weak scale becomes relatively light,
and almost compressed mass spectra for the gaugino sector can be
realized in this scenario, which is very interesting in scope of the LHC.
\end{abstract}

\maketitle


\section{Introduction}
The experimental data suggests that the see-saw scale is much lower than
the Planck scale or even the GUT scale. It is therefore natural to think 
the scale is related to the breaking of some symmetry. 
The simplest symmetry is the $B-L$ symmetry. 
In principle, the $B-L$ symmetry can be a global or a local symmetry. 
If we take it to be a global symmetry, its spontaneous breaking leads to the pseudo Nambu-Goldstone
boson, majoron. Since several experiments give severe constraints on the majoron,
it is natural to make it local gauge symmetry if we consider a higher ranked GUT such as
${\rm SO}(10)$.
The spontaneous breaking of $B-L$ symmetry can be exploited by developing
the vacuum expectation value (VEV) of a scalar multiplet $\Delta_1$
which carries $B-L=-2$. For the anomaly cancellation and to keep the low-energy
supersymmetry, its counterpart $\Delta_2$ that has $B-L=+2$ has to be included
into a theory. After the spontaneous breaking of this $B-L$ symmetry, it leads to
a massive gauge boson, $Z_{B-L}$.

In this paper we think about the possibility to break ${\rm U}(1)_{\rm B-L}$ symmetry through
the radiative corrections to the soft mass squared which is responsible for the VEV of
the ${\rm U}(1)_{\rm B-L}$ breaking in analogous to the case of RESB in the MSSM.
Here we explore such a possibility by considering the renormalization group equations (RGEs)
of the soft mass terms for the $B-L$ breaking sector.
Our resultant $B-L$ breaking scale is found to be around $v_{B-L} \simeq10^{5}$ GeV, that is
in a  sense quite appealing if we consider to incorporate the thermal leptogenesis scenario
in SUSY models because the gravitino problem put a severe constraint
on the reheating temperature as $T_R \lesssim 10^6$ GeV
for the gravitino mass of order $m_{3/2} \lesssim 100$ GeV.
 
Once we incorporate the ${\rm U}(1)_{\rm B-L}$ gauge symmetry in SUSY models,
an extra ${\rm U}(1)$ gaugino, $\tilde{Z}_{B-L}$ appears in addition to the extra gauge boson $Z_{B-L}$. 
It has recently been noticed that if there exist such an extra gaugino, it can mediate a SUSY breaking
so as to induce the gaugino masses for each SM gauge group at the two loop level,
while the scalar soft masses are generated at the one loop level. 
The Z-prime mediated SUSY breaking is basically to use an extra ${\rm U}(1)^\prime$ vector multiplet
as a field which communicates a SUSY breaking source with the visible sector. 
This setup is much more appealing and economical than the gauge-mediated
SUSY breaking.
In this mediation mechanism, it is not necessary to introduce some additional
sector as a 'messenger field', that can be implemented into a theory just as a gauge
multiplet associated with an extra ${\rm U}(1)^\prime$ gauge symmetry.
We take such an extra ${\rm U}(1)^\prime$ as a ${\rm U}(1)_{\rm B-L}$ symmetry,
and then, we can identify the messenger scale as the scale of $B-L$ symmetry
breaking scale.

\section{Radiative B-L breaking}
 The interactions between Higgs and matter superfields are
described by the superpotential %
\begin{eqnarray}%
W &=& (Y_u)_{ij} U^c_i Q_j  H_2 + (Y_d)_{ij} D^c_i Q_i  H_1 
+ (Y_e)_{ij} E^c_i L_j H_1 
\nonumber\\
&+& \mu H_1 H_2 
+ (Y_\nu)_{ij} N^c_i L_j H_2+ f_{ij} \Delta_1 N^c_i N^c_j 
\nonumber\\
&+& \mu' \Delta_1 \Delta_2 \;,
\label{superpot}
\end{eqnarray}
where the indices $i$, $j$ run over three generations, $H_u$ and $H_d$ 
denote the up-type and down-type MSSM Higgs doublets, respectively.

After developing the VEV of the $B-L$ breaking field, 
$\left<\Delta_1 \right> = v_{B-L}$,
the right-handed neutrino obtains the Majorana mass as $M_N = f v_{B-L}$.
And it gives a light neutrino mass through the see-saw mechanism as follows:
$M_\nu = m_D M_N^{-1} m_D^T$, where $m_D = Y_\nu v~(v=174\,{\rm GeV})$ 
is the Dirac neutrino mass matrix.

The soft SUSY-breaking terms which is added to the MSSM soft mass terms are given by
\begin{eqnarray}
 -  \Delta{\cal L}_{\rm soft} 
&=& ( m^2_N)_{ij} \tilde{N}_i^{\dagger} \tilde{N}_j 
+m_{\Delta_1}^2 |\Delta_1|^2 
+ m_{\Delta_2}^2 |\Delta_2|^2
\nonumber\\\
&+&
 \left((A_{\nu})_{ij} \tilde{N}_i^{\dagger}  \tilde{\ell}_j H_u  + h.c. \right)
\nonumber\\
&+& (A_f)_{ij} \Delta_1 \tilde{N}_i \tilde{N}_j  +h.c.
\nonumber\\\
&+& \frac{1}{2} M_{\tilde{Z}_{B-L}} \tilde{Z}_{B-L}  \tilde{Z}_{B-L}  +h.c. 
\label{softterms} 
\end{eqnarray}
From Eqs. (\ref{superpot}) and (\ref{softterms}),
the scalar potential relevant for the $B-L$ breaking sector can be written as
\bea
V(\Delta_1,\Delta_2)
&=& \left( |\mu'|^2 + m_{\Delta_1}^2 \right) |\Delta_1|^2 
+ \left( |\mu'|^2 + m_{\Delta_2}^2 \right) |\Delta_2|^2 
\nonumber\\
&+& \frac{1}{2} g_{B-L}^2 \left(|\Delta_1|^2 - |\Delta_2|^2 \right)^2 \;,
\eea
where we have neglected the Yukawa coupling contributions to the scalar potential.
The VEV of the $B-L$ breaking field $\Delta_1$ is determined to be
\be
|\left< \Delta_1 \right>|^2 =
- \frac{2}{g_{B-L}^2} \left( |\mu'|^2 + m_{\Delta_1}^2 \right)  \;.
\ee

\section{Z-prime mediation of SUSY breaking}
Since all the chiral superfields in the visible sector are
charged under ${\rm U}(1)_{\rm B-L}$, so all the corresponding scalars receive  soft
mass terms at 1-loop of order
\bea
\label{eqn:scalarmass}
m^2_{\tilde{q}_i} &=& \frac{8}{9} \frac{\alpha_{B-L}}{4 \pi} M_{\tilde{Z}_{B-L}}^2
\ln\left(\frac{\Lambda_S}{M_{\tilde{Z}_{B-L}}} \right),
\nonumber\\
m^2_{\tilde{\ell}_i} &=& 8\, \frac{\alpha_{B-L}}{4 \pi} M_{\tilde{Z}_{B-L}}^2
\ln\left(\frac{\Lambda_S}{M_{\tilde{Z}_{B-L}}} \right),
\eea
where $\alpha_{B-L}=g_{B-L}^2/(4\pi)$.

The MSSM gaugino masses, however,
can only be generated at 2-loop level since they do not directly couple to the ${\rm U}(1)_{\rm B-L}$,
\bea
\label{eqn:gauginomass}
M_a
&=& 4 c_a\, \frac{\alpha_{B-L}}{4 \pi} \frac{\alpha_a}{4 \pi} M_{\tilde{Z}_{B-L}}
\ln\left(\frac{\Lambda_S}{M_{\tilde{Z}_{B-L}}} \right) \;,
\eea
where $(c_1, c_2, c_3) = (\frac{92}{15}, 4,  \frac{4}{3})$.

Since these gaugino masses are proportional to $c_a$, we expect that the
gluino will typically be lighter than the others at $\mu=M_{\tilde{Z}_{B-L}}$,
so the resultant mass spectra of the gauginos are relatively compressed than the other mediation
mechanisms.

From the discussion above, we see that the gauginos are considerably lighter
than the sfermions. Taking  $M_a \simeq 100$ GeV, we find
\begin{equation}
 M_{\tilde{Z}_{B-L}} \ln\left(\frac{\Lambda_S}{M_{\tilde{Z}_{B-L}}} \right) 
\simeq 10^4 ~\TeV
 \end{equation}
 and
\begin{equation}
{m}_{\tilde{f}} \simeq  10^{-1} M_{\tilde{Z}_{B-L}} \simeq  10^{5}~\GeV.
\end{equation}
Hence, in this scheme of Z-prime mediation, all the sfermion masses become
very heavy at around $10^5$ GeV, while the gauginos are kept at at around the weak
scale, $M_a \simeq 100$ GeV, which can in principle provide a natural candidate of the dark matter.

In our choice of parameters, the gravitino mass is given by
\be
m_{3/2} = \frac{\Lambda_S^2}{\sqrt{3} M_{\rm Pl}} = \{ 24\, {\rm keV},~2.4 \,{\rm MeV},~ 240 \,{\rm MeV}\} \;.
\ee
for $\Lambda_S = \{10^7,\, 10^8, \,10^9\}$ GeV.
Hence the gravity mediation contribution to the gaugino masses is much suppressed,
and is well negligible compared to the Z-prime mediated contribution.

\section{Numerical evaluations}
Now we consider the RGEs and analyze
the running of the scalar masses $m_{\Delta_1}^2$ and
$m_{\Delta_2}^2$. The key point for implementing the radiative $B-L$
symmetry breaking is that the scalar potential $V(\Delta_1,\Delta_2)$
receives substantial radiative corrections. In particular, a
negative (mass)$^2$ would trigger the $B-L$ symmetry breaking. 
We argue that the masses of Higgs fields $\Delta_1$ and
$\Delta_2$ run differently in the way that $m^2_{\Delta_1}$ can be
negative whereas $m^2_{\Delta_2}$ remains positive.

In the numerical analysis we take input all the soft SUSY breaking parameters
to be zero at the SUSY breaking scale, in which the SUSY breaking scale
is varied in the range, $\Lambda_S = 10^7 - 10^9$ GeV,
\be
\tilde{A}_A = 0, ~ m_{\tilde{f}} = 0,~M_a = 0
\ee
and use the following inputs
\be
M_{\tilde{Z}_{B-L}}= 8.7 \times 10^5 \,{\rm GeV}\;, ~ f = 4,\ 5,\ 6, \ 7, ~g_{B-L} = 0.5 \;.
\ee
Note that $\tilde{Z}_{B-L}$ has to be decoupled at the mass scale $M_{\tilde{Z}_{B-L}}$.

\begin{figure}[h]
\includegraphics[width=.8\linewidth]{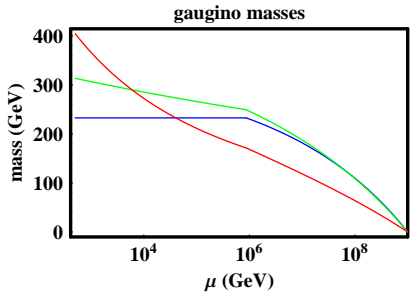}
\caption{
The evolution of the gaugino masses from the SUSY breaking scale 
to the $B-L$ breaking scale. 
The red line shows the running of the gluino mass, the green line is
the running of the ${\rm SU}(2)$ gaugino mass, and the blue corresponds to the running
of the ${\rm U}(1)_Y$ gaugino mass.
}
\label{Fig1}
\vspace{1cm}
\end{figure}

Using these inputs, in Fig.~{\ref{Fig1}}, we plot the evolution of the gaugino
masses $M_{1,2,3}$ from the SUSY breaking scale to the weak scale. 
In this plot, we fixed the SUSY breaking scale as 
$\Lambda_S = 10^9$ GeV.
It is very interesting that the gluino at the ${\tilde{Z}}_{B-L}$ scale is given as the lightest gaugino,
that is very different from most of the other models of SUSY breaking mediation.
For that reason, the gluino at the weak scale becomes relatively light,
and almost compressed mass spectra for the gaugino sector can be
realized in this scenario, which is very interesting in scope of the LHC.

\begin{figure}[h]
\includegraphics[width=.8\linewidth]{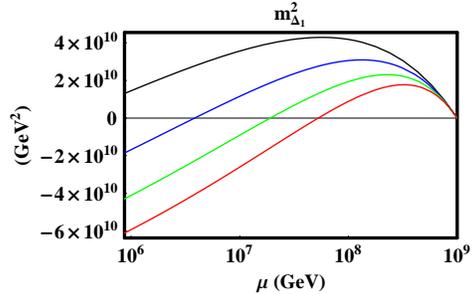}
\caption{
The evolution of the soft mass squared for the field $\Delta_1$
from the SUSY breaking scale to the $B-L$ breaking scale.
In this plot, we take the SUSY breaking scale as $\Lambda_S = 10^9$ GeV.
From top to the bottom curves, we varied the value of $f$ as $f=4,\ 5,\ 6,\ 7$.
}
\label{Fig4}
\vspace{1cm}
\end{figure}

The evolutions of the soft mass squared for the field $\Delta_1$
is plotted in Fig.~{\ref{Fig4}} for a given SUSY breaking scale as $\Lambda_S = 10^9$ GeV. 
In Fig.~{\ref{Fig4}}, from top to the bottom curves, we varied the value of $f$ as $f=4,\, 5,\, 6,\, 7$.
For example, for the case of $f=5$, the soft mass squared for the fields $\Delta_1$
goes across the zeros at the scale $10^5$ GeV toward negative value, that is nothing but the realization
of the radiative symmetry breaking of ${\rm U}(1)_{B-L}$ gauge symmetry.
The running behavior in Fig.~{\ref{Fig4}} can be understood in the following way.
At first, starting from the high energy scale, the soft mass squared increases
because of the gauge coupling contributions, and decrease of the mass squared
is caused by the Yukawa coupling that dominate over the gauge coupling contribution
at some scale. 
Next, since at the mass scale of $\tilde{Z}_{B-L}$, it is decoupled from the RGEs, 
there are only the Yukawa coupling contributions to
the soft mass squared which rapidly decreases to across the zeros.
Therefore, the radiative $B-L$ symmetry breaking can naturally be realized.

The see-saw scale, which is found to be at $v_{B-L} = 10^5$ GeV, hence
the right-handed neutrino obtains a mass of $M_N = f v_{B-L} = 5 \times 10^5$ GeV.
This scale of the right-handed neutrino is nice for the thermal leptogenesis to be viable
in supersymmetric models with gravity mediation.


\section{Summary}
We have shown that a mechanism of radiative $B-L$ symmetry breaking
can work in analogous to the RESB.
The breaking scale of $B-L$ symmetry is related to the neutrino masses
through the see-saw mechanism.
Once we incorporate the ${\rm U}(1)_{\rm B-L}$ gauge symmetry in SUSY models,
the ${\rm U}(1)_{\rm B-L}$ gaugino, $\tilde{Z}_{B-L}$ can provide all the soft masses
in the MSSM.
Then we find a link between the neutrino mass (more precisly the see-saw or $B-L$ scale
of order $10^{5}$ GeV) and the Z-prime mediated SUSY breaking scale.
In this scheme of Z-prime mediation, all the sfermion masses become
very heavy at around $10^5$ GeV, while the gauginos are kept at at around the weak
scale, $M_a \simeq 100$ GeV.
It is also very interesting that the gluino at $\tilde{Z}_{B-L}$ scale is given as the lightest gaugino,
that is very different from most of the other models of SUSY breaking mediation.
For that reason, the gluino at the weak scale becomes relatively light,
and almost compressed mass spectra for the gaugino sector can be
realized in this scenario, which is very interesting in scope of the LHC.

\section*{Acknowledgments}
We would like to thank the organizers for providing me
with an oppotunity to talk at the conference.
T. Kikuchi would like to thank K.S. Babu
 for his hospitality at Oklahoma State University.
The work of T.Kikuchi is supported by the Research
Fellowship of the Japan Society for the Promotion of Science (\#1911329).

\end{document}